\documentclass[conference]{IEEEtran}
\IEEEoverridecommandlockouts
\usepackage[top=0.75in, bottom=1.0in, left=0.625in, right=0.625in]{geometry}

\usepackage[utf8]{inputenc}
\usepackage{cite}
\usepackage{amsmath,amssymb,amsfonts}
\usepackage{algorithm}
\usepackage{algorithmic}
\usepackage{textcomp}
\usepackage{xcolor}
\usepackage{siunitx}
\usepackage{booktabs}
\usepackage{multirow}
\usepackage{colortbl}
\usepackage{tikz}
\usepackage{enumitem}
\usepackage{graphicx}
\usepackage{caption}
\usepackage{soul}

\usepackage{makecell}      
\usepackage{array}         
\usepackage{fancyhdr}
\fancypagestyle{firststyle}{\fancyhf{}
\fancyhead[L]{X. Wang, M. Ying, H. Chen, G. Qian, X. Liu, P. Ma, D. Shakya, C. Argyropoulos, and T. S. Rappaport, ``Robust Hybrid Beamforming with Liquid Crystal Antennas and Liquid Neural Networks," to appear in \textit{IEEE 103rd Vehicular Technology Conference (VTC2026-Spring),} Nice, France, Jun. 2026, pp. 1--6.}
	}
\usepackage{times}

\DeclareSIUnit{\dBm}{dBm}
\DeclareSIUnit{\dBi}{dBi}  

\def\BibTeX{{\rm B\kern-.05em{\sc i\kern-.025em b}\kern-.08em
    T\kern-.1667em\lower.7ex\hbox{E}\kern-.125emX}}

\begin{document}
\bstctlcite{BSTcontrol}
\title{\huge Robust Hybrid Beamforming with Liquid Crystal Antennas and Liquid Neural Networks
\thanks{This paper was supported by the NYU WIRELESS Industrial Affiliates Program, NYU Tandon ECE PhD Fellowship, and NSF Grant No. 2234123. The authors thank Prof. Sundeep Rangan for his valuable discussions and suggestions.}
}

\author{\IEEEauthorblockN{
Xinquan Wang$^{1}$\IEEEauthorrefmark{1}, 
Mingjun Ying$^{1}$, 
Hongren Chen$^{2}$, 
Guanyue Qian$^{1}$, 
Xingchen Liu$^{1}$, 
Peijie Ma$^{1}$, 
Dipankar Shakya$^{1}$,\\
Christos Argyropoulos$^{2}$, 
and Theodore S. Rappaport$^{1}$\IEEEauthorrefmark{2}}
\IEEEauthorblockA{$^{1}$NYU WIRELESS, New York University, Brooklyn, NY 11201, USA}
\IEEEauthorblockA{$^{2}$Pennsylvania State University, University Park, PA 16802, USA}
\IEEEauthorblockA{\{xinquanwang\IEEEauthorrefmark{1}, tsr\IEEEauthorrefmark{2}\}@nyu.edu}
}

\maketitle
\thispagestyle{firststyle}

\begin{abstract}

Sub-terahertz (sub-THz) multi-user multiple-input multiple-output (MU-MIMO) systems unlock immense bandwidth for 6G wireless communications. However, practical deployment of wireless systems in sub-THz bands faces critical challenges such as increased atmospheric absorption, reduced channel coherence time due to increased Doppler spread at higher carrier frequencies, and hardware bottlenecks as low-loss sub-THz phase shifters are difficult to realize. To overcome the hardware and channel estimation challenges of sub-THz systems, this paper proposes a hybrid beamforming (BF) framework that integrates reconfigurable liquid crystal (LC) antennas with a liquid neural network (LNN) for transmitter. Specifically, we employ an LC antenna as the analog BF stage of a hybrid BF architecture, exploiting its voltage-driven permittivity tunability to achieve high-gain beam steering without the need for lossy phase shifters. For digital BF, we utilize an ordinary differential equations-defined LNN to learn temporal channel dynamics, and use a manifold optimization technique to compress the search space. We validated the proposed method on simulated site-specific 108 GHz ray-tracing channels in an urban scenario using NYURay, a ray-tracing simulator validated against 142 GHz propagation measurements \cite{xing2021millimeter,Ju2023icc}. The 108 GHz carrier frequency matches the operating band of the LC antenna hardware. The proposed method achieves an 88.6\% spectral efficiency (SE) gain and higher robustness to imperfect channel estimation compared to the learning-aided gradient descent and gated recurrent unit machine learning baselines \cite{LAGD}, and 1.9 times higher SE than the 3GPP TR~38.901 standard antenna model \cite{3gpp2018tr38901}, highlighting the potential of LC-based hardware for sub-THz communications.

\end{abstract}

\vspace{-4mm}

\section{Introduction}
\par
Sixth-generation (6G) wireless systems are expected to deliver substantially higher data rates, motivating the use of sub-terahertz (sub-THz) bands that offer vast bandwidth for high-throughput links and sensing~\cite{rappaport2019above100ghz,bazzi2025isac,whitepaper}. However, wireless systems operating at sub-THz bands also encounter fundamental challenges, including higher atmospheric and material absorption losses, highly directional propagation characteristics, and an extreme sensitivity of the channel to user location and blockage \cite{rappaport2017overview,xing2021millimeter,scis,jiang2024thzreview}. At sub-THz carrier frequencies, the shorter wavelength increases the Doppler spread for a given user velocity, thereby reducing the channel coherence time and degrading channel estimation accuracy. Moreover, devices face a critical hardware bottleneck above 100 GHz where low-loss per-element tuning components for phase control are difficult to realize, creating a need for sub-THz capable antenna and beamforming (BF) algorithm designs that this paper aims to fulfill. The improvement of phase shifters is also an open problem that this paper aim to solve.

\par

The sub-THz hardware bottleneck of per-element phase control above 100 GHz is a major limitation in the development of scalable and potentially low-loss analog beam steering devices. Several sub-THz front-ends have been explored to form high-gain steerable beams with limited radiofrequency (RF) chains. Conventional phased arrays offer fine steering but suffer from wideband beam squint \cite{spacialwb,cai2017beamformingcodebookcompensationbeam,xing142ghz}. For further hardware reduction, lens antenna arrays exploit spatial focusing for beam selection \cite{zeng2017lens}, while dynamic metasurface antennas integrate reconfigurable combining to lower the RF-chain count \cite{wang2021dma}. Beam combining at the receiver can also reduce outage \cite{sun2014icc}. However, when frequent beam reconfiguration is required across different locations and link states, these solutions rely on lossy and costly control networks or offer only coarse beams; moreover, the associated calibration overhead can slow updates and cause alignment loss.
LC-based antennas offer a promising alternative to conventional antennas by providing continuous, voltage-driven permittivity tuning without semiconductor phase shifters, thereby enabling potentially low-loss beam steering at sub-THz frequencies. However, LC antennas introduce additional BF challenges: (i) the LC dielectric response time, on the order of milliseconds, limits the beam update rate relative to electronic phase shifters, (ii) the achievable phase tuning range per element constrains the effective steering aperture, and (iii) mutual coupling between adjacent LC unit cells complicates independent per-element control and necessitates carefully designed codebooks~\cite{Wang2023LCMetasurface, Wang2024LCMetasurface}.

\par
While scalable hardware enables efficient beam steering, practical sub-THz deployment also demands BF algorithms that can adapt to rapid channel variations and remain robust under degraded channel estimation \cite{robust_network,telemom}.
Several BF methods have therefore been proposed \cite{jayaprakasam2017,yang2020bayesian,wgan,dynthz}.
While iterative methods employing angle-error statistics \cite{jayaprakasam2017} or Bayesian inference \cite{yang2020bayesian} improve robustness of BF to rapid channel variations, they suffer from high computational overhead. Alternatively, neural network-based methods like \cite{dynthz} reduce latency by utilizing fusion-separation models to approximate optimal BF, but they remain limited in their ability to capture the continuous physical dynamics of changing channels.
Designed to effectively learn real-world dynamic systems, a continuous-time architecture governed by ordinary differential equations (ODEs) named liquid neural network (LNN) was introduced \cite{hasani2022closed}. By learning the ODE parameters to mimic the time-varying behavior of biological synapses, LNNs demonstrate exceptional robustness in processing noisy, continuous-time data.

\par
This paper proposes a hybrid BF method that is robust to imperfect channel estimation, using a reconfigurable LC antenna and an LNN. Specifically, we utilize an LC antenna array capable of continuous dielectric tuning, enabling electronic beam steering at 108 GHz in the sub-THz band, a frequency within the 105--108 GHz operating range of the LC antenna hardware.
The proposed hybrid architecture operates in two sequential stages: an analog BF first selects an LC antenna radiation pattern from a discrete codebook, and a digital BF then applies baseband precoding to manage multi-user interference.
The framework formulates analog BF as a discrete LC antenna pattern selection problem from a pre-optimized codebook of 19 radiation patterns. For digital BF, we utilize an LNN to optimize the digital BF matrix via manifold optimization following \cite{rwmmse}. Unlike prior work that relies on statistical channel models, we validate the proposed method in a realistic urban scenario using simulated site-specific 108 GHz channels generated by NYURay \cite{ying2026site,Kanhere2023icc,kanhere2024calibration, ying2025location}, a ray-tracing simulator calibrated against 142 GHz measurements \cite{xing2021millimeter,Ju2023icc,xing142ghz,nie2013indoor,ju142ghz}; the 108 GHz carrier frequency was selected to match the operating band of the LC antenna hardware.
Simulation results demonstrate that our approach outperforms the learning-aided gradient descent (LAGD) and gated recurrent unit (GRU) baselines \cite{LAGD} by 88.6\% in SE and shows increased robustness to imperfect channel estimation against degraded channel estimation. Furthermore, the LC antenna yields 1.9 times higher SE compared to the 3GPP TR~38.901 standard antenna model \cite{3gpp2018tr38901}.
\par
The remainder of this paper is organized as follows: Section II introduces the downlink sub-THz MU-MIMO hybrid BF system model and defines the hybrid BF optimization problem, Section III describes the  LC antenna design, analog BF design and the LNN-based digital BF workflow, Section IV reports numerical results, and Section V concludes the paper.
\par
\textit{Notation}: We denote scalar, vector, matrix, transpose, Hermitian, inverse, Frobenius norm, Hadamard product, expectation, trace, complex Gaussian distribution and identity matrix as $a,\mathbf{a},\mathbf{A},(\cdot)^{\mathrm{T}},(\cdot)^{\mathrm{H}},(\cdot)^{-1},\|\cdot\|_F,\odot, \mathbb{E}(\cdot), \det(\mathbf{\cdot}),\mathrm{Tr}(\mathbf{\cdot}), \mathcal{CN}(\cdot,\cdot),\mathbf{ I}$, respectively. 

\section{System Model and Problem Formulation}
\subsection{System Model}
Consider a downlink sub-THz multi-user MIMO (MU-MIMO) hybrid BF communication system where a base station (BS) with $M$ transmit antenna elements simultaneously serves $K$ users, each of which are equipped with $N_k$ receive antennas, where $M \gg N_k$. Let $s_k \in \mathbb{C}$ denote the data symbol intended for user $k$, where $s_k \sim \mathcal{CN}(0,1)$.
The unit-variance Gaussian assumption models the transmitted data symbols for capacity analysis \cite{tse2005fundamentals}; the actual multipath channel is captured by the site-specific ray-tracing channel matrix $\mathbf{H}_k^{(p)}$ in \eqref{eq:channel_multipath} without assuming Rayleigh fading.

The proposed hybrid architecture uses the LC antenna array as the analog front-end at the BS, replacing conventional phase-shifter networks. Each of the $M$ LC antenna elements connects to a dedicated RF chain feeding a digital baseband processor. The two BF stages operate sequentially: the LC pattern is first selected to maximize aggregate channel gain, and then the digital precoder $\mathbf{W}$ is computed to suppress inter-user interference. This paper focuses on the transmit-side hybrid BF design at the BS; receive-side processing at the UEs is beyond the scope of this work, as the BS-side BF dominates performance in massive MIMO systems where $M \gg N_k$.

For the digital BF part, we denote $\mathbf{w}_k \in \mathbb{C}^{M\times 1}$ as the BF vector for user $k$.
Therefore, the BS transmits the signal
$\mathbf{u} = \sum_{k=1}^{K} \mathbf{w}_k s_k$.
For analog BF, we utilize LC antennas discussed in Section \ref{sec:lcant}. Every LC antenna here is embedded with $n_p$ predefined antenna patterns that steering at $n_p$ different directions. Let $p \in \{1, 2, \ldots, n_p\}$ denote the index of the antenna pattern. When LC antenna pattern $p$ is selected, the channel matrix between the base station and user $k$ is denoted as $\mathbf{H}_k^{(p)} \in \mathbb{C}^{N_k \times M}$. The superscript $(p)$ indicates that the channel matrix depends on the selected antenna pattern, because different patterns produce different radiation gains in different directions. The additive Gaussian noise vector as $\mathbf{n}_k\in \mathbb{C}^{N_k \times 1}$ with distribution $\mathcal{CN}(\mathbf{0},\sigma^2\mathbf{I})$, where $\sigma^2$ is the variance of the noise.
Therefore, the received signal $\mathbf{y}_k$ at user $k$ is
\begin{equation}\label{eq:hybrid_received}
\mathbf{y}_k =
{\mathbf{H}_k^{(p)} \mathbf{w}_k s_k}+
{\sum_{j \neq k} \mathbf{H}_k^{(p)} \mathbf{w}_j s_j}+
\mathbf{n}_k,
\end{equation}
where the $s_k$ and $\mathbf{n}_k$ are assumed to be independent for different $k$.
For clarity, we define $N\triangleq \sum_{k=1}^{K}N_k$ as the total number of receive antennas, and denote $\mathbf{y}\triangleq [\mathbf{y}_1^\mathrm{T}, \mathbf{y}_2^\mathrm{T},\cdots,\mathbf{y}_K^\mathrm{T}]^\mathrm{T} \in \mathbb{C}^{N\times 1}$, $\mathbf{W}\triangleq [\mathbf{w}_1, \mathbf{w}_2,\cdots,\mathbf{w}_K] \in \mathbb{C}^{M\times K}$, $\mathbf{s}\triangleq [s_1, s_2,\cdots,s_K]^\mathrm{T} \in \mathbb{C}^{K\times 1}$, $\mathbf{n}\triangleq [\mathbf{n}_1^\mathrm{T}, \mathbf{n}_2^\mathrm{T},\cdots,\mathbf{n}_K^\mathrm{T}]^\mathrm{T} \in \mathbb{C}^{N\times 1}$, and 
$\mathbf{H}^{(p)}\triangleq [(\mathbf{H}_1^{(p)})^\mathrm{T}, (\mathbf{H}_2^{(p)})^\mathrm{T},\cdots,(\mathbf{H}_K^{(p)})^\mathrm{T}]^\mathrm{T} \in \mathbb{C}^{N\times M}$.
Therefore, \eqref{eq:hybrid_received} can be rewritten as $\mathbf{y} = \mathbf{H}^{(p)}\mathbf{W}\mathbf{s}+\mathbf{n}.$

\subsection{Channel Model}
The channel matrix $\mathbf{H}_k^{(p)}$ is generated using NYURay ray-tracing simulation at a carrier frequency of \SI{108}{\giga\hertz}~\cite{ying2026site,Kanhere2023icc,kanhere2024calibration}. The ray-tracing identifies $L$ propagation paths between the base station and user $k$. Each path $\ell$ has a departure direction $(\theta_\ell^t, \phi_\ell^t)$ from the transmitter and an arrival direction $(\theta_\ell^r, \phi_\ell^r)$ at the receiver, where $\theta$ denotes the elevation angle and $\phi$ denotes the azimuth angle.
The channel matrix is the sum of contributions from all $L$ paths:
\begin{equation}\label{eq:channel_multipath}
\mathbf{H}_k^{(p)} = \sum_{\ell=1}^{L} \alpha_\ell \, G^{(p)}(\theta_\ell^t, \phi_\ell^t) \, \mathbf{a}_r(\theta_\ell^r, \phi_\ell^r) \, \mathbf{a}_t^H(\theta_\ell^t, \phi_\ell^t).
\end{equation}
The terms in equation~\eqref{eq:channel_multipath} are defined as follows. The complex path gain $\alpha_\ell = |a_\ell| e^{-j2\pi f_c \tau_\ell}$ includes the amplitude $|a_\ell|$ and the phase shift due to propagation delay $\tau_\ell$, where $f_c$ is the carrier frequency. The term $G^{(p)}(\theta_\ell^t, \phi_\ell^t)$ is the antenna gain of LC pattern $p$ in the departure direction of path $\ell$. 
The vector $\mathbf{a}_t(\theta, \phi) \in \mathbb{C}^{M \times 1}$ is the transmit array response vector, and $\mathbf{a}_r(\theta, \phi) \in \mathbb{C}^{N_k \times 1}$ is the receive array response vector.
For a linear array with element spacing $d$, the array response vector is
$\mathbf{a}(\theta, \phi) = \left[1, \, e^{j (2\pi d / \lambda) \sin\theta\cos\phi}, \, \ldots, \, e^{j(2\pi d / \lambda)(W-1)\sin\theta\cos\phi}\right]^\mathrm{T},$
where $\lambda$ is the wavelength. For the transmit array, $d = d_t$ and $W = M$. For the receive array, $d = d_r$ and $W = N_k$.

\subsection{Problem Formulation}\label{sec:prob}
In practice, the BS does not have access to accurate $\mathbf{H}^{(p)}$ (the stacked channel matrix of all $K$ users, as defined above).
Instead, all optimization is performed based on a noisy channel estimate.
In order to evaluate the robustness of the proposed algorithms against imperfect channel estimation, we add estimation errors into the channel matrices generated from \eqref{eq:channel_multipath} for simulation. We denote the estimated channel for user $k$ as $\hat{\mathbf{H}}_k^{(p)}=\mathbf{H}_k^{(p)}+\mathbf{E}_k^{(p)}$, and the stacked estimated channel as $\hat{\mathbf{H}}^{(p)}=[(\hat{\mathbf{H}}_1^{(p)})^\mathrm{T},\ldots,(\hat{\mathbf{H}}_K^{(p)})^\mathrm{T}]^\mathrm{T}$. We define the channel estimation error (CEE) as
  $  \mathrm{CEE}\triangleq 10\log_{10}\left( \mathbb{E}[\|\mathbf{E}^{(p)}\|_F^2] / \mathbb{E}[\|\mathbf{H}^{(p)}\|_F^2] \right)$.

Since the BS only observes $\hat{\mathbf{H}}^{(p)}$, i.e., the noisy estimate of the stacked channel matrix, all digital BF computation should be carried out using $\hat{\mathbf{H}}^{(p)}$ instead of the true channel $\mathbf{H}^{(p)}$. Therefore, the optimization objective function for the digital BF is
\begin{equation}\label{SE}
    R= \sum_{k=1}^{K}\log_2 \det (\mathbf{I}+ \gamma_k)
\end{equation}
where $\gamma_k$ denotes the signal-to-interference-plus-noise ratio matrix of the user $k$, given by
\begin{equation}\label{rkinv}
    \gamma_k= \hat {\mathbf{H}}^{(p)}_k\mathbf{w}_k(\hat {\mathbf{H}}^{(p)}_k\mathbf{w}_k)^{\mathrm{H}}
        \left(\sum_{j\neq k}^{K} \hat {\mathbf{H}}^{(p)}_k\mathbf{w}_j(\hat {\mathbf{H}}^{(p)}_k\mathbf{w}_j)^{\mathrm{H}}+\sigma^2\mathbf{I}\right)^{-1}.
\end{equation}

\par
The basic problem is to maximize $R$ by finding an optimal $p$ and $\mathbf{W}$ with a sum power constraint, given as $\mathrm{Tr}(\mathbf{WW}^{\mathrm{H}})\leq P$, where $P$ is the sum power budget of the transmit antenna. The hybrid BF optimization problem can be formulated as
\begin{equation}\label{problem}
        \max_{\mathbf{W},p}\ \ R, \ \ \ 
        \mathrm{s.t.}\ \mathrm{Tr}(\mathbf{WW}^{\mathrm{H}})\leq P,\ p\in \{1, 2, \ldots, n_p\}
\end{equation}

\begin{figure}[t]
  \centering
  \includegraphics[page=1,width=\linewidth]{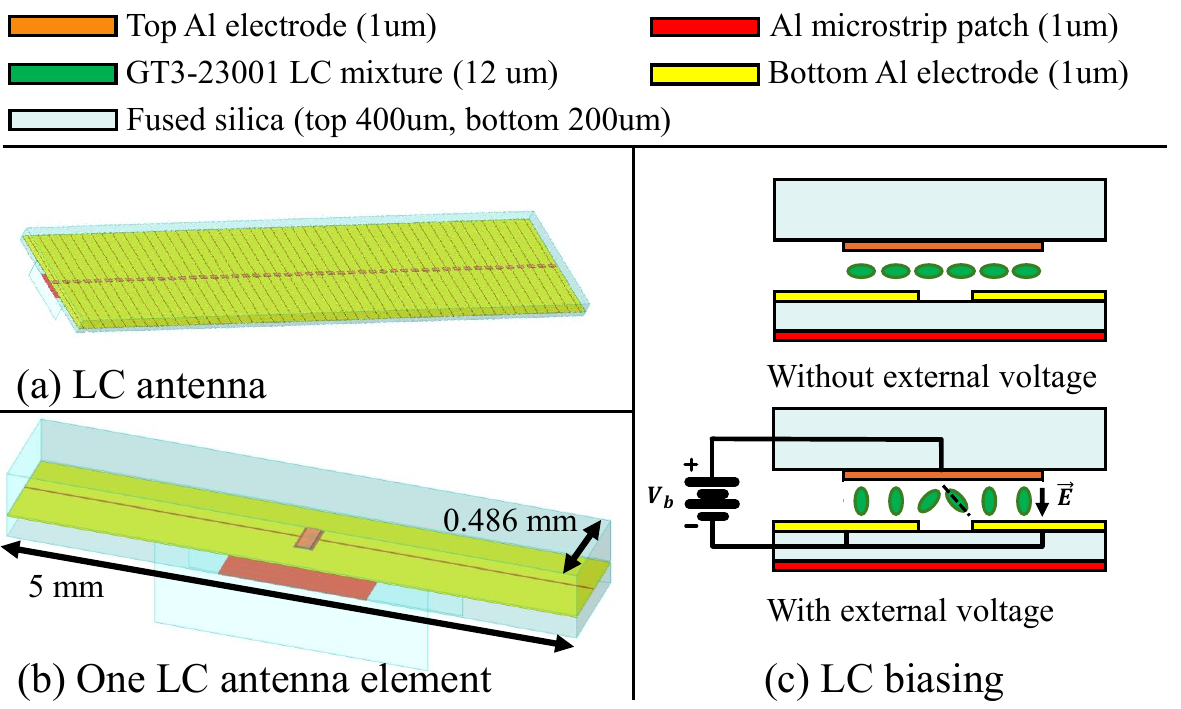}
  \caption{Schematic of the LC antenna used. (a) Model of the LC-based antenna structure consisting of 48 unit cells. (b) Model of a single LC unit cell integrated into the ground plane (yellow) of a widened microstrip feed line (red). (c) Lateral cross-sectional schematic of a single unit cell showing the LC cavity, biasing electrodes, and the electric-field-induced reorientation of the LC molecules, which modifies the effective permittivity within the cavity \cite{Wang2023LCMetasurface}.
  }
  \label{fig:dma_beamforming_pdf}
\end{figure}

\section{Robust Hybrid Beamforming with LNN and LC Antenna}
The proposed framework decouples hybrid BF into two sequential stages. In the first stage, analog BF selects an LC antenna radiation pattern $p$ from a discrete codebook to maximize the effective channel gain. In the second stage, digital BF computes the precoding matrix $\mathbf{W}$ in baseband using the effective channel corresponding to the selected pattern. Since the analog stage operates in the RF domain via LC voltage control and the digital stage operates in baseband, the two stages do not interfere with each other \cite{shu2018ana}.

\subsection{Liquid Crystal Antenna and Analog Beamforming}\label{sec:lcant}

To realize scalable analog beam steering at sub-THz frequencies, we utilize a LC-based reconfigurable antenna structure \cite{Wang2024LCMetasurface,Wang2023LCMetasurface}, as illustrated in Fig.~\ref{fig:dma_beamforming_pdf}.
Unlike conventional phased arrays dependent on phase shifters, the proposed architecture comprises a 48-element linear array
, where each unit cell incorporates GT3-23001 LC material. By dynamically controlling the bias voltage of individual elements, the antenna synthesizes directional beams via holographic BF in the sub-THz band.
This mechanism enables agile steering across a \SI{90}{\degree} field of view with a \SI{5}{\degree} beamwidth in the 105--108 GHz band, achieving a \SI{6.87}{dB} gain per element without requiring phase shifters, thereby reducing power waste in the RF chain~\cite{ying2024wastefactor}.

The discrete codebook consists of $n_p=19$ pre-optimized radiation patterns covering steering angles from $-45^{\circ}$ to $+45^{\circ}$ in $5^{\circ}$ increments. Each pattern is obtained by applying the bias-voltage distribution generated by the holographic algorithm across the 48 LC-loaded elements in full-wave electromagnetic simulation, targeting maximum directional gain at the corresponding steering angle while suppressing sidelobes. The resulting set of $n_p$ voltage configurations and the associated radiation patterns $\{G^{(p)}(\theta,\phi)\}_{p=1}^{n_p}$ constitute the analog BF codebook.

Selecting state $p\in\{1,\ldots,n_p\}$ produces a radiation pattern with directional gain $G^{(p)}(\theta,\phi)$, which directly weights each path contribution in \eqref{eq:channel_multipath}.

\subsection{Liquid Neural Network}
LNNs are ODE-based recurrent neural networks that model dynamical systems without separate numerical solvers \cite{hasani2022closed}. We leverage LNNs to capture the temporal dynamics of the channel and BF process.
Consider a basic first-order ODE $\mathrm{d}x(t)/\mathrm{d}t=-x(t)+S(t)$, where $S(t)=f(i(t))(a-x(t))$ represents the synaptic current between neurons, and $x(t), a$ represent the current state and the bias.
Substituting $S(t)=f(i(t))(a-x(t))$ into $\mathrm{d}x(t)/\mathrm{d}t=-x(t)+S(t)$ yields a linear ODE in $x(t)$:
\begin{equation}
\frac{\mathrm{d}x(t)}{\mathrm{d}t}=-(1+f(i(t)))\,x(t)+a\,f(i(t)).
\end{equation}
Using the integrating-factor solution of linear ODEs, we obtain the integral closed-form
that can be simplified and vectorized to
\begin{equation}
        \mathbf{x}(t)=(\mathbf{x}(0)-\mathbf{a})\odot e^{-\mathbf{o}_\tau t-\int_0^t\mathbf{f}(\mathbf{i}(s),\theta_{\mathbf{f}})\mathrm{d}s}+\mathbf{a},
\end{equation}
where $\mathbf{x}(t)\in \mathbb{R}^{D\times 1},\mathbf{i}(t)\in \mathbb{R}^{C\times 1},\mathbf{o}_\tau\in \mathbb{R}^{D\times 1},\mathbf{a}\in \mathbb{R}^{D\times 1},\mathbf{f}$ are the hidden state, external inputs with $C$ features, time constant parameter, bias, and the function of the neural network with parameters $\theta_{\mathbf{f}}$,
respectively. 
We follow the approximation in \cite{hasani2022closed} and remove the integral term, resulting in
\begin{equation}\label{ltc}
\mathbf{x}(t)\approx \mathbf{b}\odot
e^{-[\mathbf{o}_\tau+\mathbf{f}(\mathbf{i}(t),\theta_{\mathbf{f}})]t}
\odot \mathbf{f}(-\mathbf{i}(t),\theta_{\mathbf{f}})
+\mathbf{a},
\end{equation}
which retains the ODE-driven temporal evolution while avoiding separate numerical solvers.
Since the exponential decay in \eqref{ltc} drives $\mathbf{x}(t)$ rapidly towards $\mathbf{a}$, the exponential term is replaced with sigmoid gates.
Finally, to make model more adaptive to changing scenarios, we treat $\mathbf{a}$ and $\mathbf{b}$ as learnable by parameterizing them with NN heads $\mathbf{g}$ and $\mathbf{h}$, respectively.
To better capture temporal dependencies, the input to each liquid neuron consists of both the current input vector $\mathbf{i}$ and the previous hidden state $\mathbf{x}$.
Accordingly, the closed-form continuous-time model is written as
\begin{equation}\label{cfc}
\begin{split}
\mathbf{x}(t)=&\,\sigma\!\big(-\mathbf{f}(\mathbf{x},\mathbf{i};\theta_{\mathbf{f}})t\big)\odot
\mathbf{g}(\mathbf{x},\mathbf{i};\theta_{\mathbf{g}})\\
&+\Big[\mathbf{1}-\sigma\!\big(-\mathbf{f}(\mathbf{x},\mathbf{i};\theta_{\mathbf{f}})t\big)\Big]\odot
\mathbf{h}(\mathbf{x},\mathbf{i};\theta_{\mathbf{h}}),
\end{split}
\end{equation}
where $\theta_{\mathbf{g}}$ and $\theta_{\mathbf{h}}$ denote the parameters of NN heads $\mathbf{g}$ and $\mathbf{h}$, respectively.

The LNN architecture is well-suited for the BF optimization in \eqref{problem} for two reasons. First, the sub-THz channel evolves continuously in time due to UE mobility and environmental dynamics, and the ODE-based state evolution in \eqref{cfc} provides a natural inductive bias for modeling such continuous-time dynamics, unlike discrete-time architectures (like GRUs) that process channel snapshots independently \cite{lnn_tracking,wang2024liquid}. Second, the sigmoid gating mechanism in \eqref{cfc} bounds the hidden state trajectory, acting as an implicit regularizer that improves robustness when channel estimates are noisy \cite{welai,ztecom}.

\subsection{Workflow of the Proposed Method}

\subsubsection{Manifold Optimization}
Leveraging the characteristic that $M \gg N$ in massive MIMO systems, we adopt a manifold optimization approach \cite{rwmmse} to reduce the search space of the optimization problem in \eqref{problem}. We model $\mathbf{E}$ Gaussian as \cite{hoydis2013massive}, and the randomness from the Gaussian noise prevent rank deficiency, making $\hat{\mathbf{H}}^{(p)}$ full row rank.
Under the manifold optimization framework, the optimal $\mathbf{W}$ for the BS can be projected onto the row space of $\hat{\mathbf{H}}^{(p)}$ as
$\mathbf{W}=\mathbf{\hat H}^{\mathrm{H}}\mathbf{X}$,
allowing the proposed LNN to optimize a base matrix $\mathbf{X}\in\mathbb{C}^{N\times K}$ rather than optimize $\mathbf{W}$ directly.
When $M\gg N$, manifold optimization compresses the search space from $\mathbb{C}^{M\times K}$ to $\mathbb{C}^{N\times K}$, significantly reducing the number of parameters the LNN must optimize \cite{gmml,wang2023energyefficient}.

\subsubsection{Forward Propagation}
In the forward propagation, the LNN updates the digital BF matrix using the estimated channel $\hat{\mathbf H}$ as input.  
Since the elements in $\mathbf{\hat{H}}$ are extremely small, potentially leading to numerical underflow. Therefore, we normalize the channels by $\mathbf{\hat{H}}_{\mathrm{n}}=\mathbf{\hat H}/\sigma$, and feed $\mathbf{\hat{H}}_{\mathrm{n}}$ into three layers of liquid neurons that are fully connected. For the simulations with analog BF, we input all $\mathbf{\hat{H}_{k,\mathrm{n}}}$ into the neural network, and obtain $\mathbf{X}=\mathrm{LNN}(\mathbf{\hat{H}}_{\mathrm{n}})$. Finally, considering the power constraints in \eqref{problem}, we can compute the $\mathbf{W}$ as
\begin{equation}\label{pwrcst}
    \mathbf{W}=\sqrt{P/{\mathrm{Tr}(\mathbf{\mathbf{\hat{H}}}^{\mathrm{H}} \mathbf{X} (\mathbf{\hat{H}}^{\mathrm{H}} \mathbf{X})^{\mathrm{H}})}}\cdot \mathbf{\hat{H}}^{\mathrm{H}} \mathbf{X},
\end{equation}

\subsubsection{Backward Propagation}
To train the proposed LNN, we employ a logarithmic loss to encourage a fair SE distribution among users. This design prevents the network from collapsing to a single-user solution and improves training stability.
Specifically, given the per-user SE $R_1,\ldots,R_K$, the log loss is defined as
$
\mathcal{L}
= -\sum_{k=1}^{K}\log \left(\max (\epsilon, R_k)\right),
$
where $\epsilon$ is a small positive constant to avoid numerical instability.
For simulations with analog BF, we only select the loss from $p$ with the best SE as calculated in \eqref{SE}.
Then, we use an Adam optimizer to optimize the parameters, as
$
    \theta^*=\theta+\alpha \cdot \mathrm{Adam(\nabla_\theta \mathcal{L},\theta}),
$
where $\theta,\theta^*$ are NN parameters before and after update, respectively, and $\alpha$ is the learning rate.

\section{Numerical Results}
\begin{table}[t]
\centering
\caption{Simulation Parameters}
\label{tab:simulation_parameters}
\footnotesize
\renewcommand{\arraystretch}{1.15}
\resizebox{0.48\textwidth}{!}{%
\begin{tabular}{ll}
\toprule
\textbf{Parameter} & \textbf{Value} \\
\midrule
\multicolumn{2}{l}{\textit{NYURay} \cite{ying2025nyu,ying2026site} \textit{Ray-Tracing  Configuration}} \\
Carrier frequency & \SI{108}{\giga\hertz} \\
Number of rays & $10^6$ (5 max reflections) \\
Propagation modes & Reflection, penetration, diffraction \\
\midrule
\multicolumn{2}{l}{\textit{Antenna Configuration}} \\
BS antenna ($M$) & $48 \times 1$ (LC antenna elements) \\
UE antenna ($N_k$)& $4 \times 1$ ULA (Isotropic, \SI{0}{\dBi}) \\
Polarization & Vertical \\
\midrule
\multicolumn{2}{l}{\textit{Deployment Scenario}} \\
Location & Brooklyn MetroTech Commons, NYC \\
Geographic boundaries & 40.688°--40.699°N, 73.972°--73.992°W \\
Coverage area & \SI{1.8}{\kilo\meter} $\times$ \SI{1.2}{\kilo\meter} \\
(BS height, UE height) & (20 m, 1.5 m) \\
\bottomrule
\end{tabular}%
}
\end{table}
\begin{figure}[t]
  \centering
  \includegraphics[page=1,width=0.8\linewidth]{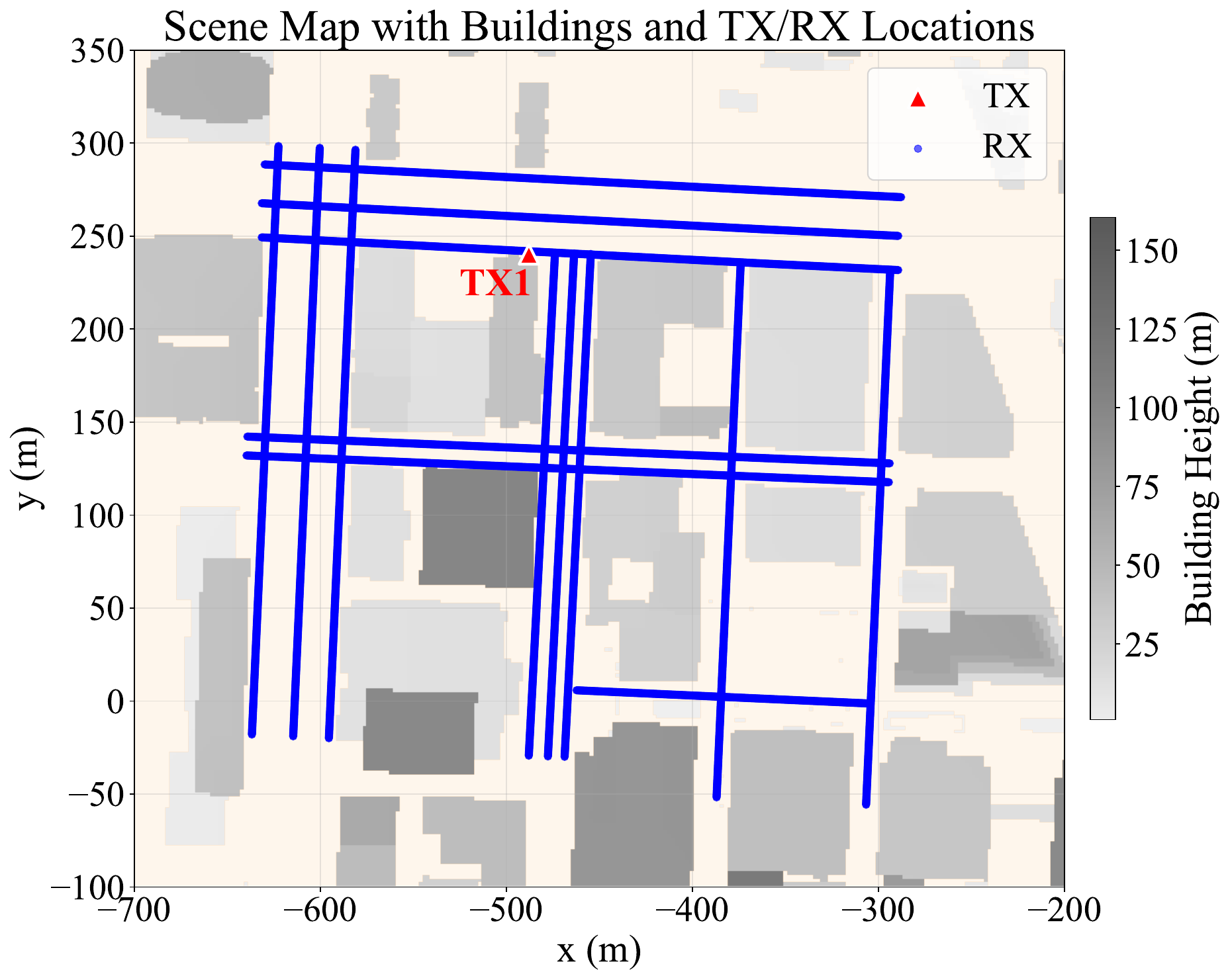}
  \caption{Site-specific simulation scenario in Brooklyn MetroTech Commons, Brooklyn, NY, generated using NYURay at 108 GHz. TX1 marks the BS equipped with a $48 \times 1$ LC antenna array at 20 m height with $10^{\circ}$ downtilt. The blue traces represent UE routes along streets at 1.5 m height, to emulate mobile users traversing the coverage area. For each UE position, ray-tracing channels were generated to evaluate the hybrid BF performance under realistic urban propagation conditions.}
  \vspace{-8pt}
  \label{fig:scene}
\end{figure}

\subsection{Scenario and Baselines}
We evaluated the proposed system using a site-specific simulation of the Brooklyn MetroTech Commons, Brooklyn, NY, a typical urban scenario covering \SI{1.8}{km} $\times$ \SI{1.2}{km} at \SI{108}{GHz}, as shown in Fig. \ref{fig:scene}~\cite{shakya2025umi}. We place the BS at a typical urban environment \cite{3gpp2018tr38901}, which is \SI{20}{m} above the ground. UEs operate at street level (\SI{1.5}{m} height) positioned at realistic locations. NYURay~\cite{ying2025nyu, ying2026horama} generated channels using $10^6$ rays with up to five reflections. Table~\ref{tab:simulation_parameters} provides complete scene parameters.

We use $K=4,P=30\mathrm{\ dBm},\alpha=0.01$ and run simulations on an NVIDIA RTX 4090 GPU using PyTorch 2.5.1. For baselines, we consider LAGD \cite{LAGD} and a GRU-based model. To emphasize the effect of LC antennas, we compare all three methods using 3GPP TR~38.901 antennas \cite{3gpp2018tr38901}.

\subsection{Evaluation on Beamforming Performance}
\begin{figure}[t]
  \centering
\includegraphics[page=1,width=0.8\linewidth]{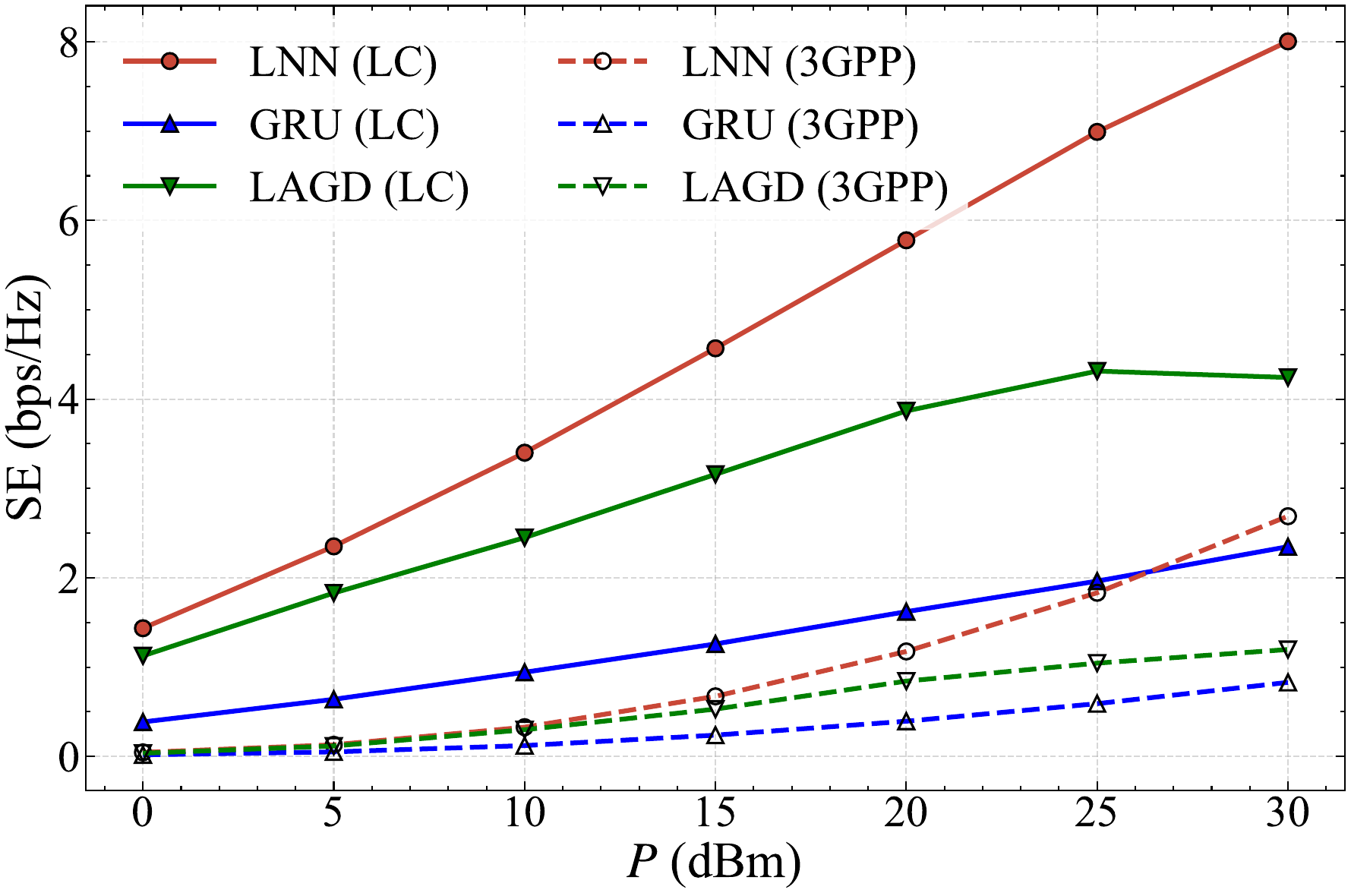}
\caption{SE vs. BS total transmit power budget $P$ at CEE $= -10$ dB. The LNN with LC antennas achieves 88.6\% higher SE than LAGD at $P=30$ dBm. All LC antenna configurations outperform their 3GPP counterparts by over 1.9 times in SE.}
  \vspace{-3mm}
  \label{fig:rate_vs_power}
\end{figure}

In Fig. \ref{fig:rate_vs_power}, we present the spectral efficiency (SE) of the proposed method, under conditions of fixed CEE as -10 dB and varying $P$.
The results show that the SE of all the algorithms increases with an increasing $P$, and the proposed method outperforms all the baselines, being 88.6\% higher than the LAGD when $P$ is 30 dBm.
LNN outperforms baselines due to its ability to learn channel dynamics via its ODE architecture.
Besides that, a significant SE gain (over 1.9 times) is observed comparing the LC antenna and 3GPP antennas, showcasing the benefit of using LC-based hardware and coherent analog BF algorithm in sub-THz frequencies.

\subsection{Evaluation on Robustness to Inaccurate Channel Estimation}

\begin{figure}[t]
  \centering
\includegraphics[page=1,width=0.8\linewidth]{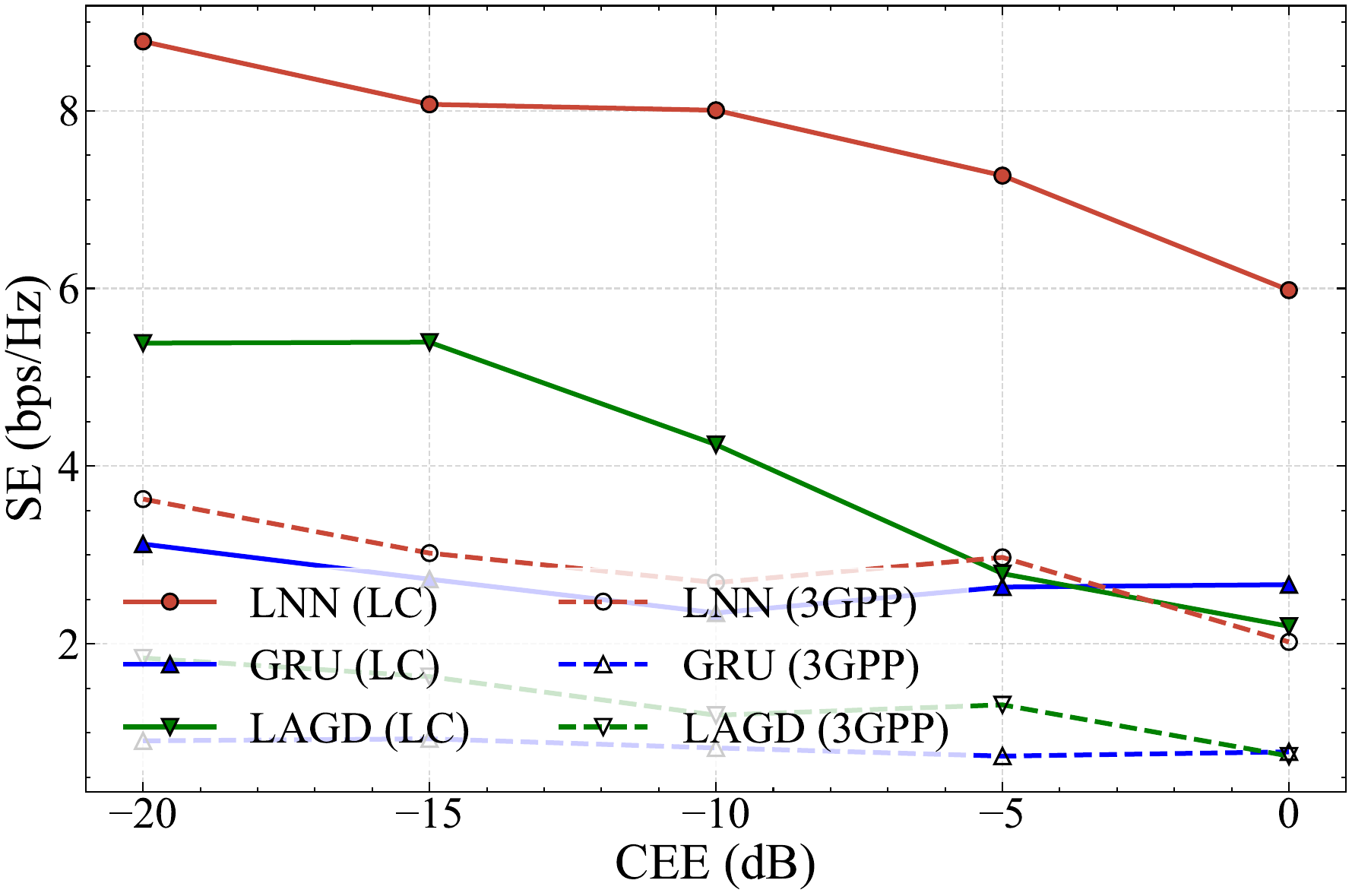}
  \caption{Robustness evaluation SE vs. channel estimation error (CEE) at $P=30$ dBm. The proposed LNN approach demonstrates higher robustness to imperfect CSI compared to the LAGD and GRU baselines.}
  \vspace{-10pt}
  \label{fig:csi_robustness}
\end{figure}

In Fig. \ref{fig:csi_robustness}, we fix $P= 30$ dBm and evaluate the performance of the algorithms as the CEE increases from -20 dB to 0 dB. 
As shown in the figure, the LNN (LC) demonstrates significantly stronger robustness against high CEE compared to LAGD (LC). Specifically, the LNN (LC) experiences only a 31.7\% reduction in SE (from 8.8 to 6.0 bps/Hz), whereas LAGD (LC) suffers a 55.4\% reduction (from 5.4 to 2.4 bps/Hz).

The stronger robustness of the LNN stems from two architectural properties. First, the sigmoid gating in \eqref{cfc} inherently bounds the hidden state updates, preventing large estimation errors from causing unbounded changes in the BF output. In contrast, LAGD uses gradient-based iterative updates that can amplify noise in the channel estimate across iterations. In addition to gating, the manifold projection $\mathbf{W}=\hat{\mathbf{H}}^{\mathrm{H}}\mathbf{X}$ constrains the precoder to lie in the row space of the estimated channel, acting as implicit regularization that keeps the precoder structurally aligned with the dominant channel subspace. A full robustness characterization including per-user SE variance analysis and SE cumulative distribution under varying CEE levels is deferred to future work.

\section{Conclusion}
A hybrid BF framework was presented for sub-THz MU-MIMO systems, combining a reconfigurable LC antenna for analog BF with an ODE-inspired LNN for digital BF, validated on simulated ray-tracing channels at 108 GHz in an urban scenario.
The 88.6\% SE gain of the LNN over the LAGD and GRU baselines is attributed to two factors: (i) the sigmoid gating mechanism in the LNN bounds hidden-state updates, preventing noise amplification that affects the iterative gradient steps in LAGD, and (ii) the continuous-time ODE formulation provides an inductive bias better matched to the temporal evolution of sub-THz channels than the discrete-time GRU architecture. The LNN also demonstrated stronger robustness to imperfect channel estimation, with only a 31.7\% SE reduction as CEE increased from $-$20 dB to 0 dB, compared to 55.4\% for LAGD.
Separately, the LC antenna achieved 1.9 times higher SE than the 3GPP TR~38.901 antenna array. This gain is attributed to the higher per-element directivity of the LC design, where each of the 48 elements achieves 6.87 dB gain through BF, and codebook-based pattern selection that concentrates radiated energy toward users with a 5-degree beamwidth. A controlled comparison with matched array aperture and element count between the LC and 3GPP arrays is left to future work.
Future work also includes varying the number of users and antennas, per-user SE distribution analysis, and experimental validation of the LC antenna with field measurements.

\bibliographystyle{IEEEtran}
\bibliography{references}

\end{document}